\documentclass[11pt,twoside]{article}
\usepackage{./asp2014}

\aspSuppressVolSlug
\resetcounters

\begin{document}

\title{Massive and Evolved Stars with the ngVLA}
\author{Thomas J. Maccarone,$^1$ Saida Caballero-Nieves$^2$, Nathan Smith$^3$,  Nora L\"utzgendorf$^4$
\affil{$^1$Texas Tech University, Lubbock, TX, USA; \email{Thomas.Maccarone@ttu.edu}}
\affil{$^2$Florida Institute of Technology, Melbourne, FL, USA}
\affil{$^3$University of Arizona, Tucson, AZ, USA}
\affil{$^4$Space Telescope Science Institute, Baltimore, MD, USA}
}

\paperauthor{Sample~Author1}{Author1Email@email.edu}{ORCID_Or_Blank}{Author1 Institution}{Author1 Department}{City}{State/Province}{Postal Code}{Country}
\paperauthor{Sample~Author2}{Author2Email@email.edu}{ORCID_Or_Blank}{Author2 Institution}{Author2 Department}{City}{State/Province}{Postal Code}{Country}
\paperauthor{Sample~Author3}{Author3Email@email.edu}{ORCID_Or_Blank}{Author3 Institution}{Author3 Department}{City}{State/Province}{Postal Code}{Country}

\begin{abstract}

The Next Generation Very Large Array will have excellent sensitivity
for detecting the thermal emission from massive stars and from red
giants.  This will allow direct imaging of the winds for a large
number of hot massive stars.  It will also allow using the radio
emission for the massive stars as a way to detect stars to allow high
resolution measurements can be made, even with large extinction.  A
few examples of the utility of the high resolution measurements are
given: dynamics of globular clusters with red giants, detection of
intermediate mass stripped stars in binaries, and measurement of
masses of stars in massive binaries.

\end{abstract}

\section{Description of the problem}

Winds of hot massive stars are strong, spatially extended sources of
free-free emission in the radio band.  The acceleration of stellar
winds is often probed via optical and ultraviolet emission lines, but
is rarely directly resolved.  Stellar winds from many classes of stars
allow us to have bright radio sources that can be used as dynamical
tracers, both for measuring the masses of massive stars in binaries
and for measuring the gravitational potential fields in globular
clusters, and should also provide a check on orbits around the
Galactic Center.  Estimates of the masses of the most massive stars
have been extremely difficult in the past, with luminosities and
single star evolutionary tracks and/or the Eddington limit used for
mass estimates, rather than dynamical measurements typically used as a
technique.  By providing a tool for obtaining precise astrometry and
high dynamic range, even in highly obscurred parts of the sky, radio
measurements of stellar masses will provide crucial information for
understanding massive stars.

With globular clusters, it is quite controversial about whether the
clusters contain intermediate mass black holes.  Radial velocity
measurements and proper motions of stars in the optical band are a
valuable tool for probing the inner dynamics of star clusters, but are
complicated by hard-to-constrain crowding effects.  Using radio
tracers of stellar dynamics will allow a complementary test of whether
there are intermediate mass black holes in clusters, and will also
allow measurements of the proper motions of the most extincted
clusters.

Finally, observations with the Next Generation Very Large Array
(ngVLA) should also be able to resolve stellar winds directly, and
provide direct imaging constraints on the clumpiness and acceleration
of stellar winds.  They should also be able, through direct imaging,
to determine whether stellar winds are anisotropic, as may be expected
for rapidly rotating stars.

\section{Scientific importance and Astronomical Impact}

\subsection{Understanding massive star winds}

The winds from massive stars affect the whole evolution of the stars
as well as the properties of the supernovae and massive remnants
produced later in their lifetimes (see e.g \citet{Smith14} for a review).
Understanding the stellar wind mass loss rates, and how the winds are
accelerated, is thus an important process for interpreting supernovae,
understanding the formation of compact objects, and understanding the
deposition of metals into the Universe.  Furthermore, the kinetic
power of the winds themselves can sometimes be important for powering
superbubbles.

Direct imaging of stellar winds has been done for some low mass stars,
but never for massive stars.  First, one can study, in detail, the
spatial scale on which clumping in stellar winds takes place.  It can
be well established from the difference between line strengths in
P~Cygni profiles, which scale linearly with density, and free-free and
recombination line tracers, which scale as density squared, that
significant clumping must be present \citep{Abbott81,Smith14}.  For
massive stars, the clumping factors are typically $\sim5$
\citep{Repolust04,Markova04}. Analysis which ignores clumping
overestimates the mass loss rates by factors of $\sim3$ in most cases,
but in some extreme cases, overestimates of a factor of 10 have been
suggested \citep{Fullerton06}.  Similar clumping factors have been
suggested for cool evolved stars\citep{Harper10}.

Theoretical considerations suggest that the clumping is taking place
primarily in the inner regions of the stellar wind, but at the present
time, there is no direct imaging evidence of this.  The spatial scale
for the clumping should be smaller than the size scale of the star,
but with the Long Baseline Major Option, this may be resolvable with
ngVLA.  Regardless, it should be straightforward to determine whether
the clumping is taking place preferentially in the inner region of the
wind.

Additionally, studies can be made of the departures from isotropy of
stellar winds.  Be star winds show a variety of lines of evidence for
being disk-like.  Understanding these stars' winds is essential for
developing a universal theory of mass loss, since it is generally
believed that stellar winds from rapidly rotating stars are launched
through a fundamentally different mechanism than those from other
stars\citep{Araya18}.

The core figure of merit for stellar wind observations would be the
ability to make good measurements of the flux density per beam in each
beam in reasonable exposure times.  As a conservative flux estimate,
we start with the equations from \citep{Wright75} for a
constant velocity, smooth stellar wind, using the form from \citet{Gudel02}:

\begin{equation}
R_{thick} = 8\times10^{28} \left(\frac{\dot{M}}{v}\right)^{2/3} T^{-0.45} \nu^{-0.7}
\end{equation}

where $R_{thick}$ is the radius of the optically thick part of the
stellar wind, $\dot{M}$ is the mass loss rate in $M_\odot$ yr$^{-1}$, $T$ is the wind temperature in $K$, $v$ is the wind velocity in
km/sec, and $\nu$ is the frequency in Hz at which the observations are
made.

Then, the flux density, $S_\nu$ in mJy from the stellar wind is given by:
\begin{equation}
S_\nu = 9\times10^{10} \left(\frac{\dot{M}}{v}\right)^{4/3} T^{0.1} d_{pc}^{-2} \nu^{0.6}
\end{equation}
if $R_{thick}$ is greater than the stellar radius and
\begin{equation}
S_\nu = 5\times10^{39} \left(\frac{\dot{M}}{v}\right)^{2} T^{-0.35} R_*^{-1} d_{pc}^{-2}\nu^{-0.1}
\end{equation}
if $R_{thick}$ is smaller than the stellar radius, $R_*$, with $d$ the distance to the star in pc.

For a massive star with a reasonably high mass loss rate, we can take
$M_\odot = 10^{-5} M_\odot$ yr $^{-1}$, $v$=1000 km/sec, $T=10^4 K$
and $\nu$=15 GHz.  We find that the wind will be optically thick out
to about 30 AU, considerably bigger than the stellar radius of massive
stars.  Next we can take $d=1$ kpc, and we find that $S_\nu$ = 6 mJy
for such a star, with an angular size scale of 30 milliarcseconds.
With 1000 km baselines, the angular resolution at 15~GHz would be 4
milliarcseconds, allowing for about 50 beams across the source.  If
one assumes homogeneous surface brightness, this would yield 100
$\mu$Jy/beam flux densities, meaning that $\sim100\sigma$ detections
would be possible in each beam.  Having so many well-detected beams
would allow for careful studies of source structure, source
variability, and source spectra, by making similar measurements at
other frequencies.  This would thus allow, in turn, for many stars, the
decomposition of the radio emission into components from the thermal
part of the wind, synchrotron emission from shocks (which should
mostly contribute at lower frequencies), and from the stellar
photosphere (which should contribute most at higher frequencies ---
see the contribution by Chris Carilli in this volume).

\subsection{Understanding massive star masses and orbits}

It is expected that the most massive stars should be found
predominantly in binary (or other multiple) star systems with other
massive stars, unless they have undergone mergers \citep{Sana12}.
Still, with the most massive stars, binary companions can be emitting
a small fraction of the total starlight in the system, and hence can
be extremely difficult to detect.  Additionally, the radial velocity
wobbles in such systems can also be very small.  Precise astrometric
measurements of the stars can provide an alternative means of
establishing binarity in many cases, and, in the best cases, of
estimating the masses of the stars.

The highest mass stars are expected to be about $\sim300 M_\odot$ or
more \citep{Crowther10,Crowther16}, and they show mass loss rates of
$2-5 \times10^{-5} M_\odot$/yr, despite being in the Large Magellanic
Cloud, where low metallicity might be expected to suppress mass
loss\citep{Crowther10}.  A small number of similar stars have been
seen in the Milky Way, but extinction has generally made it difficult
to identify such stars in our own Galaxy; the Magellanic Cloud stars
themselves obviously cannot be measured with the ngVLA, but these
similar stars can be identified in Galactic radio plane surveys, and
studied carefully with additional follow-up.  Conservatively assuming
a typical 10 kpc distance, along with $\dot{M}$ = $2\times10^{-5}
M_\odot$/yr, $v_W$ of 2500 km/sec, we can expect radio flux densities
of about 50 $\mu$Jy for these stars at 15 GHz, and angular sizes of
about 3 milliarcseconds (and perhaps a bit smaller if the winds are
significantly hotter).

These objects are thus well-suited for astrometric work with the
ngVLA.  Positional measurements with precision of about $50\mu$arcsec
should be reached in about one hour of observations, and with a set of
longer observations, it should be possible to make geometric parallax
measurements for these stars.  Furthermore, if there are binary
companions with even 10\% of the flux of the primary stars, they
should be detectable as well.  In such a case, following the proper
motions of both stars gives a ``visual binary'' type mass estimate for
the two stars.

To establish that a system is a binary without detecting its companion
star, one needs to make a measurement of the acceleration of the
motion of the object.  As a rule-of-thumb, let us consider a binary
with separation of 100~AU (a fairly typical value in star-forming
regions -- \citet{Griffiths18}).  Such a binary, if the total system
mass is 100 $M_\odot$ will have an orbital period of about 100 years.
If the two stars are 10 $M_\odot$ and $90 M_\odot$, then the more
massive star will have a reflex motion of about 10 AU, which will
correspond to a total variation in position of 1 milliarcsecond.  Over
the course of 10 years of observations, it will move through
12$^\circ$ of its orbit, meaning that the deviation from a pure
straight line will be about 20 $\mu$arcsec.  The $\approx 0.1 AU$/year
velocity of the star's motion, corresponding to 0.5 km/sec, is beyond
present capabilities for radial velocity precision for massive stars,
and is likely to remain that way -- thus even if such a star can be
detected in the infrared or optical bands with high signal-to-noise,
its binarity can be studied only astrometrically.   For a star like the
one discussed in the preceding paragraph, spending about 20 hours per
year on obtaining positions accurate to 10 $\mu$arcsec each would
yield a data set sufficient for ruling out a straight-line fit.  For
more nearby star-forming regions, of course, such measurements could
be made more quickly.  For supergiant stars where the stellar
photosphere may exceed the brightness of the stellar wind, such
measurements could be made even more easily using the photospheric
emission.

\subsection{Detecting intermediate mass stripped stars}

Stripped stars produced by binary stellar evolution are likely to be
quite common in the Galaxy due to binary interactions, and to be an
important source of ionizing radiation, due to their exceptionally hot
temperatures \citep{Goetberg18}.  Despite their importance,
they can be difficult to detect when they have a massive binary
companion, because of the higher luminosities (especially in the less
extincted red bands) of their companion stars.  The stellar winds from
these stars can make them detectable in the radio.  Taking numbers
typical for these stars \citep{Vink17}, $\dot{M}$=$10^{-7} M_\odot$/yr,
$v=2000$ km/sec, $T$=50000 K, $d=2$ kpc, and an observing frequency of
30 GHz, the expected radius of the optically thick part of the stellar
wind is 0.2 AU (meaning that these objects would appear to be point
sources with inverted radio spectra unless they are very nearby), and
the expected flux density is $0.2 \mu$Jy.  In such systems, then,
either two objects will be seen as a resolved binary, or the stellar
winds are likely to be interacting such that synchrotron emission will
be seen, and the systems will be identifiable as colliding wind
binaries.

\subsection{Proper motions of globular cluster and Galactic Center stars}

The search for intermediate mass black holes in globular clusters has
been a long-running source of controversy.  A few approaches are
typically used -- searches for accretion (e.g. \citet{Bahcall75};
\citet{Maccarone04}; chapter by Wrobel in this volume), searches for
radial velocity excesses toward the center of the cluster
\citep{noyola08}, and searches for proper motion velocity excesses
toward the center of the cluster \citep{Watkins15}.  All of these
approaches have caveats. E.g. searches for evidence of accretion have
been upper limits in current work, and only in a few cases is the gas
content of the cluster well-established.  With optical measurements,
there are always potential problems associated with crowding in the
inner regions of the clusters.  Crowding problem will, in fact, be
{\it worse} with JWST than with Hubble in the modes with large fields
of view, but can be ameliorated with aperture masking.  For radial
velocities there can be additional complications due to the effects of
rotation and binarity of the stars measured.  Radio-based proper
motions are ideal for solving this problem, so long as a sufficiently
abundant set of dynamical tracers can be detected in the radio band.

Detection of photospheric emission from globular cluster stars in the
radio will require prohibitive exposure times.  Attempts to make radio
measurements of globular cluster star motions thus rely on the use of
radio-bright tracers.  At the present time, pulsars and X-ray binaries
may provide such tracers for a subset of clusters.  These can be
useful for bulk proper motion measurements and, in M4, for geometric
parallax measurements.  No cluster has a large enough number of
detected pulsars and X-ray binaries to make useful radial velocity
dispersion measurements.

It may become possible with ngVLA to make sure measurements with
pulsars in Terzan~5, but ideally, we would be able to obtain larger
samples of objects which are bright enough in radio to allow proper
motion measurements.  In fact, the stellar winds from red giants
should provide such tracers.  For a wind mass loss rate of $10^{-8}
M_\odot$/yr \citep{Meszaros08}, a wind speed of 10 km/sec, and a
distance of 4 kpc, the expected radio flux density is 11 $\mu$Jy,
while the flux density would be 4 $\mu$Jy for $2\times10^{-9}
M_\odot$/yr \citep{Cohen76}, and clumping in the winds may make the radio
fluxes brighter than these values.

To this sample of objects we can add the millisecond pulsars and X-ray
binaries, which will generally be significantly brighter than $11
\mu$Jy.  With such a flux, using the Long Baseline Major Option, using
only the antennas on the long baselines, with angular resolution of
about 1.5 milliarcsecond (approximately the same as the expected
angular size of the optically thick part of the stellar wind), each
one-hour measurement will have a positional accuracy of about 150
$\mu$arcseconds.  Over a 2-year time baseline, then, the proper motion
accuracy of each star would be obtained to a precision of about 0.8
km/sec.  As a result, then, the precision of the velocity dispersion
within a bin would be set by the sampling of the velocity
distribution, rather than by the precision of the measurements, unless
very large numbers of stars were detected.

The uncertainty on the one-dimensional velocity dispersion from a set
of Gaussian-distributed measurements will be $\sigma_v \sqrt{1/(2N)}$,
where $\sigma_v$ is the velocity dispersion of the sample and $N$ is
the number of stars measured.  With two dimensions measured, there
will be an extra $\sqrt{2}$ term, meaning that $\sigma_{v,m}$ for a
bin, the measured velocity dispersion within that bin, will have a
precision of $\frac{1}{2}\sqrt{1/N}$.  Thus, with 20 stars per bin,
30\% changes in the velocity dispersion will be measurable at the
$3\sigma$ level.  The massive globular clusters in the Galaxy tend to
contain a few hundred to a few thousand bright red giant stars,
meaning that the velocity dispersions should be measurable to
$\sim5\%$ for these clusters, providing significantly better tests of
the IMBH hypothesis than can currently be achieved.  For some of the
less massive clusters it will, at least, be possible to use the few
brightest red giants to measure the clusters proper motions, and often
geometric parallaxes, even for the case of heavily extincted clusters
where Gaia is not effective.

\section{Connection to unique ngVLA capabilities}

The interest in resolved emission at high frequencies on angular
scales of $\sim10-100$ milliarcseconds is fundamentally ngVLA science.
At the present time, only eMERLIN covers a similar angular resolution
scale, but eMERLIN has much worse sensitivity than the ngVLA will, is
located far North so that a much smaller fraction of the Galactic
Plane is observable, and is in a location where the prevailing weather
conditions make an expansion beyond 22 GHz unlikely to occur (or to
provide much useful data).  For the astrometric work on massive stars,
ngVLA allows seeing through large extinction and looking at the
Galactic massive star population in a way that cannot be done easily
in optical and infrared.  For the astrometric work on giants in
globular clusters, the ngVLA measurement allow avoidance of the
crowding and the highest possible angular resolution through the use
of the Long Baseline Major Option.  Without the Long Baseline Major
Option, the proper motion precision could still be obtained by
achieving higher signal-to-noise measurements, but it would be more
difficult to ensure that there was no crowding affecting the
measurements.

\acknowledgements We thank Selma de Mink for suggestions which helped
start this project, and Jay Strader, Eva Noyola and Laura Watkins for
discussions of globular cluster dynamics.



\end{document}